\documentclass[12pt, centerh1]{article}
\usepackage{colonequals,color,multirow,natbib}
 \usepackage{graphicx}

\usepackage{amsfonts, amsmath, amssymb, marvosym}
\usepackage{algorithm}
\usepackage{bm,relsize}
\usepackage{geometry}
\usepackage{lscape}
\usepackage{mathtools}
\usepackage{caption}
\usepackage{theorem}
\usepackage{placeins}
\usepackage{hyperref}
\usepackage{enumerate}
\usepackage{float}
\usepackage{caption}
\usepackage{subcaption}

\newcommand{\be}{\begin{equation}}
\newcommand{\ee}{\end{equation}} 
\newcommand{\beq}{\begin{equation*}}
\newcommand{\eeq}{\end{equation*}}

\newcommand{\diag}{\,\mbox{diag}}
\newcommand{\sign}{\,\mbox{sign}}

\newcommand{\tr}{\,\mbox{tr}}
\newcommand{\load}{\mathbf\Lambda}

\newcommand{\vecA}{\mathbf{A}}

\newcommand{\vecD}{\mathbf{D}}
\newcommand{\vecY}{\mathbf{Y}}
\newcommand{\vecy}{\mathbf{y}}
\newcommand{\vecx}{\mathbf{x}}
\newcommand{\vecw}{\mathbf{w}}
\newcommand{\vecW}{\mathbf{W}}
\newcommand{\vecX}{\mathbf{X}}
\newcommand{\vecU}{\mathbf{U}}

\newcommand{\vecV}{\mathbf{V}}
\newcommand{\vecz}{\mathbf{z}}

\newcommand{\vecN}{\mathbf{N}}

\newcommand{\vecOmeg}{\mbox{\boldmath$\Omega$}}

\newcommand{\varthet}{\mbox{\boldmath$\vartheta$}}
\newcommand{\vectheta}{\mbox{\boldmath$\theta$}}

\newcommand{\vecmu}{\mbox{\boldmath$\mu$}}
\newcommand{\vecbeta}{\mbox{\boldmath$\beta$}}
\newcommand{\vecalpha}{\mbox{\boldmath$\alpha$}}

\newcommand{\vecDelta}{\mbox{\boldmath$\Delta$}}
\newcommand{\mSigma}{\mbox{\boldmath$\Sigma$}}
\newcommand{\matsig}{\mSigma}

\newcommand{\lsum}{\sum_{i=1}^n}

\newcommand{\gsum}{\sum_{g=1}^G}

\newcommand{\zig}{\hat{z}_{ig}}

\theorembodyfont{\upshape}

\begin{document}

\title{Unsupervised Learning via Mixtures of Skewed Distributions with Hypercube Contours}

\author{Brian C.\ Franczak \qquad\qquad \ \ Cristina Tortora\\ Ryan~P.~Browne \qquad\qquad Paul D.\ McNicholas\thanks{Department of Mathematics \& Statistics, University of Guelph, Guelph, Ontario, N1G 2W1, Canada. E-mail: paul.mcnicholas@uoguelph.ca.}% <-this % stops a space
}
\date{}%Department of Mathematics \& Statistics, University of Guelph.}

\maketitle

\begin{abstract}
Mixture models whose components have skewed hypercube contours are developed via a generalization of the multivariate shifted asymmetric Laplace density. Specifically, we develop mixtures of multiple scaled shifted asymmetric Laplace distributions. The component densities have two unique features: they include a multivariate weight function, and the marginal distributions are also asymmetric Laplace. We use these mixtures of multiple scaled shifted asymmetric Laplace distributions for clustering applications, but they could equally well be used in the supervised or semi-supervised paradigms. The expectation-maximization algorithm is used for parameter estimation and the Bayesian information criterion is used for model selection. Simulated and real data sets are used to illustrate the approach and, in some cases, to visualize the skewed hypercube structure of the components.
\end{abstract}

\smallskip
\noindent \textbf{Keywords.} Finite Mixture Models; Shifted Asymmetric Laplace; EM Algorithm; Skewed Distribution; Multiple Scaled Distributions.

\section{Introduction}\label{sec:introduction}
Unsupervised learning, or cluster analysis, can be lucidly defined as the process of sorting like-objects into groups. Finite mixture models are a convex combination of probability densities; accordingly they are natural choice for performing cluster analysis. The general finite mixture model has density
%\beq
$$f(\vecx\mid\varthet)= \sum_{g=1}^G \pi_g f_g(\vecx\mid\vectheta_g),$$
%\eeq 
where $\pi_g >0$, such that $\sum_{g=1}^G \pi_g = 1$, are the mixing proportions and $f_1(\vecx~|~\vectheta_1),\ldots,f_G(\vecx~|~\vectheta_G)$ are the component densities. To date, multivariate Gaussian component densities have been the focal point in the development of finite mixtures for clustering \citep[e.g.,][]{fraley02a,banfield93,celeux95,ghahramani97}. Their popularity can be attributed to their mathematical tractability and they continue to be prominent in clustering applications \citep[e.g.,][]{maugis09,scrucca09,punzo13}. 

Around the beginning of the $21$st century work using mixtures of multivariate-$t$ distributions began to surface \citep{peel00}, and in the last 5 years has flourished \citep[e.g.,][]{greselin10a,andrews11a,baek11}. One limitation of the multivariate-$t$ distribution is that the degrees of freedom parameter is constant across dimensions. \cite{forbes13} exploited the fact that a random variable $\vecU\in\mathbb{R}^p$ from a multivariate-$t$ distribution is a normal variance-mean mixture to give a generalized multivariate-$t$ distribution, where the degrees of freedom parameter can be uniquely estimated in each dimension of the parameter space. %, a multivariate Pearson type VII distribution \citep[see, e.g.][vol. 2 chap.\ 28]{johnson94}, a multivariate $K$ model \citep[cf.][]{eltoft06a}, and a multivariate normal inverse Gaussian distribution \citep[cf.][]{karlis09}. However, a parameter estimation procedure is only presented for the multivariate-$t$ generalizations.
Herein, we discuss the development and application of a mixture of multiple scaled shifted asymmetric Laplace (MSSAL) distributions which, unlike the multivariate-$t$ generalizations, have the ability to parameterize skewness in addition to location and scale. Furthermore, the level sets of our MSSAL density are guaranteed to be convex, making mixtures thereof ideal for clustering applications. %(e.g., Figure%s~\ref{MSContours} and

\section{Mixtures of Multiple Scaled Shifted Asymmetric \\Laplace distributions}\label{sec:MSSAL}

\subsection{Shifted Asymmetric Laplace}

\cite{kotz01} show that a random vector $\vecV$ arising from a multivariate asymmetric Laplace distribution can be generated through the relationship 
%\be\label{eq:CAL}
$\vecV=W\vecalpha + \sqrt{W}\vecN$,
%\ee
where $\vecN$ is a $p$-dimensional random vector from a multivariate Gaussian distribution with mean $\mathbf{0}$ and covariance matrix $\mSigma$, and $W$, independent of $\vecN$, is a random variable following an exponential distribution with rate 1. To facilitate cluster analysis, \cite{franczak14} introduce a $p$-dimensional shift parameter $\vecmu$ and consider a random vector $\vecX=\vecV+\vecmu$. It follows that $\vecX$ will have the stochastic representation
\be\label{eq:SAL}
\vecX = \vecmu + W\vecalpha + \sqrt{W}\vecN,
\ee
where $W$ and $\vecN$ are as previously defined. Herein, we follow \cite{franczak14} and use the notation $\vecX\backsim\mathcal{SAL}\left(\vecalpha,\mSigma,\vecmu\right)$ to mean that the random vector $\vecX$ is distributed multivariate shifted asymmetric Laplace (SAL) with skewness parameter $\vecalpha\in\mathbb{R}^p$, $p\times p$ scale matrix $\mSigma$, and location parameter $\vecmu\in\mathbb{R}^p$. 

We can see from \eqref{eq:SAL} that the random vector $\vecX\mid w$ is multivariate Gaussian with mean $\vecmu+w\vecalpha$ and scale matrix $w\mSigma$. Therefore, $\vecX$ is a normal variance-mean mixture with density
\be\label{eq:SALNMVM}
f(\vecx\mid\vecmu,\mSigma,\vectheta)=\int_{0}^{\infty}{\phi_p\left(\vecx\mid\vecmu+w\vecalpha,w\mSigma\right)h_W(w)dw},
\ee
where $\phi_p\left(\vecx\mid\vecmu+w\vecalpha,w\mSigma\right)$ is the multivariate Gaussian density with mean $\vecmu+w\vecalpha$ and covariance matrix $w\mSigma$, and $h_W(w)=e^{-1}$. Formally, the random vector $\vecX\backsim\mathcal{SAL}\left(\vecalpha,\mSigma,\vecmu\right)$ has density
\be\label{eq:SALDens}
\xi\left(\vecx\mid\vecalpha, \matsig,\vecmu\right)=
\frac{2\exp\{(\vecx-\vecmu)'\matsig^{-1}\vecalpha\}}
{(2\pi)^{p/2}\vert\matsig\vert^{1/2}}
\left(\frac{\delta\left(\vecx,\vecmu\mid\matsig\right)}
{2+\vecalpha'\matsig^{-1}\vecalpha}\right)^{\nu/2}
K_{\nu}\left(u\right),
\ee
where $\nu=(2-p)/2$, $u = \sqrt{(2+\vecalpha'\matsig^{-1}\vecalpha)\delta\left(\vecx,\vecmu\mid\matsig\right)}$, $\delta\left(\vecx,\vecmu\mid\matsig\right)$ is the squared Mahalanobis distance between $\vecx$ and $\vecmu$, $K_{\nu}$ is the modified Bessel function of the third kind with index~$\nu$, and $\vecalpha$, $\matsig$, and $\vecmu$ are previously defined. In the one-dimensional case, \eqref{eq:SALDens} reduces to
\be\label{eq:SALUDens}
\xi\left(x\mid\mu,\gamma\right)=
\frac{1}{\gamma}
\exp{\left\{-\frac{\vert x\vert}{\sigma^2}\left(\gamma-\mu\cdot\sign{(x)}\right)\right\}},
\ee
where $\mu$ is a location parameter, $\gamma = \sqrt{\alpha^2+2\sigma^2}$, $\alpha$ is a skewness parameter and $\sigma^2$ is a scale parameter \citep[cf.][]{kotz01}.

\subsection{Multiple Scaled Distributions}

\cite{forbes13} show that the density of a random variable $\vecY$ arising from a normal variance-mean mixture can be written
\be\begin{split}\label{eq:NMVM2}
f(\vecy\mid\vecmu,\vecD,\vecA,\vectheta)&=\int_{0}^{\infty}\dots\int_{0}^{\infty}\phi_p\left(\vecy\mid\vecmu,\vecD\vecA\vecDelta_\vecw\vecD'\right)\\
&\qquad\quad\times f_\vecW\left(w_1,\dots,w_p\mid\vectheta\right)dw_1\dots dw_p,
\end{split}\ee
where $\phi_p\left(\vecy\mid\vecmu,\vecD\vecA\vecDelta_\vecw\vecD'\right)$ is the multivariate Gaussian density with mean $\vecmu$ and covariance matrix $\vecD\vecA\vecDelta_\vecw\vecD'$, $\vecD$ is a matrix of eigenvectors, $\vecA$ is a diagonal matrix of eigenvalues, $\vecDelta_\vecw=\diag\left(w_1,\dots,w_p\right)$ is a diagonal weight matrix where $w_1,\dots,w_p$ are independent, and $f_\vecW\left(w_1,\dots,w_p\mid\vectheta\right) = f_W\left(w_1\mid\vectheta_1\right)\times\dots\times f_W\left(w_p\mid\vectheta_p\right)$ is a $p$-variate density function.

Notably,
\begin{align}\label{eq:NMVM3}
\phi_p\left(\vecy\mid\vecmu,\vecD\vecA\vecDelta_\vecw\vecD'\right) 
& = \prod_{j=1}^p{\phi_1\left([\vecD'\vecy]_j\mid [\vecD'\vecmu]_j, a_jw_j^{-1}\right)}\\
&= \prod_{j=1}^p{\phi_1\left([\vecD'(\vecy-\vecmu)]_j\mid0,a_jw_j^{-1} \right)}
\end{align}
and, therefore,
\be\begin{split}\label{eq:NMVM4}
f(\vecy\mid\vecmu,\vecD,\vecA,\vectheta)&=\prod_{j=1}^p\int_0^{\infty}\phi_1\left([\vecD'(\vecy-\vecmu)]_j\mid 0, a_jw_j^{-1} \right)\\
&\qquad\qquad\qquad\qquad\times f_{W_j}\left(w_j\mid\vectheta_j\right)dw_j,
\end{split}\ee
where $[\vecD'(\vecy-\vecmu)]_j$ is the $j$th element of $\vecD'(\vecy-\vecmu)$, $\phi_1\left([\vecD'(\vecy-\vecmu)]_j\mid 0, a_jw_j^{-1} \right)$ is the univariate Gaussian density with mean 0 and variance $a_jw_j^{-1}$, $f_{W_j}\left(w_j\mid\vectheta_j\right)$ is the density of an univariate random variable $W_j>0$, and $a_j$ is the $j$th eigenvalue of the matrix $\vecA$.
%, i.e., the $j$th eigenvalue. %Note: if all the weights are equal to one, a standard multivariate Gaussian distribution is recovered. 
% \citep[cf.][]{forbes13}.

Amalgamating~\eqref{eq:SALNMVM} and~\eqref{eq:NMVM2} gives an expression for the density of a generalized multivariate SAL distribution. Explicitly, this density is given by
\be\begin{split}\label{eq:SALNMVM2}
\xi(\vecx\mid\vecalpha,\vecD,\vecA,\vecmu)=&\int_{0}^{\infty}\dots\int_{0}^{\infty}\phi_p\left(\vecx\mid\vecmu+\vecDelta_\vecw\vecalpha,\vecD\vecA\vecDelta_\vecw\vecD'\right)\\
&\qquad\quad~\times h_\vecW\left(w_1,\dots,w_p\right)dw_1\dots dw_p,
\end{split}\ee
where $\phi_p\left(\vecx\mid\vecmu+\vecDelta_\vecw\vecalpha,\vecD\vecA\vecDelta_\vecw\vecD'\right)$ is the multivariate Gaussian density with mean $\vecmu+\vecDelta_\vecw\vecalpha$ and scale matrix $\vecD\vecA\vecDelta_\vecw\vecD'$, and $h_\vecW(w_1,\dots,w_p)=h(w_1)\times\dots\times h(w_p)$ is the density of a $p$-variate exponential distribution with $h(w_j)=e^{-1}$ for $j=1,\dots,p$.

To simplify the derivation of our parameter estimates we set $\vecDelta_\vecw\vecalpha = \vecOmeg\vecbeta$, where $\vecbeta\in\mathbb{R}^p$ 
%is a transformed skewness parameter 
and $\vecOmeg=\vecD\vecA\vecDelta_\vecw\vecD'$.
Given this parameterization it follows from~\eqref{eq:NMVM3} that~\eqref{eq:SALNMVM2} can be written
\begin{equation*}\begin{split}\label{eq:NVMN5}
\xi(\vecx&\mid\vecbeta,\vecD,\vecA,\vecmu)=\\
&\prod_{j=1}^p\int_0^{\infty}\phi_1\left(\left[\vecD'(\vecx-\vecmu)\right]_j\mid \left[\vecD'(\vecA\vecDelta_\vecw\vecbeta)\right]_j,a_jw_j \right)h_{W_j}\left(w_j\right)dw_j,
\end{split}\end{equation*}
where  $\left[\vecD'(\vecx-\vecmu)\right]_j$ is the $j$th element of $\vecD'(\vecx-\vecmu)$, $\phi_1\left(\left[\vecD'(\vecx-\vecmu-\vecA\vecDelta_\vecw\vecbeta)\right]_j\mid \left[\vecD'(\vecA\vecDelta_\vecw\vecbeta)\right]_j,a_jw_j\right)$ is the univariate Gaussian density with mean $\left[\vecD'(\vecA\vecDelta_\vecw\vecbeta)\right]_j$ and covariance matrix $a_jw_j$, and $h_{W_j}\left(w_j\right) = e^{-w_j}$, for $w_j>0$. 

Thus, the density of MSSAL distribution is given by
\be\begin{split}\label{eq:MSSAL}
&{h}\left(\vecx\mid\vecbeta,\vecD,\vecA,\vecmu\right)=\\
&\prod_{j=1}^p\frac{1}{\gamma_j}\exp{\left\{\frac{-\big|\vecD'\left[\vecx-\vecmu\right]_j\big|}{a_j}\left[\gamma_j-\left[\vecA\vecD'\vecbeta\right]_j\cdot\sign(\vecD'[\vecx-\vecmu]_j)\right]
\right\}},
\end{split}\ee
where $\gamma_j=\sqrt{\left[\vecA\vecD'\vecbeta\right]_j^2+2a_j},$ and $a_j$, $\vecbeta$, $\vecD$, $\vecA$, and $\vecmu$ are previously defined. It follows that the density of a mixture of MSSAL distributions is
%\be\label{eq:mixMSSAL}
$$f(\mathbf{x}\mid\varthet)=\gsum{\pi_g{h}\left(\vecx\mid\vecbeta_g,\vecD_g,\vecA_g,\vecmu_g\right)},$$
%\ee
where $\pi_g$ are the mixing proportions and ${h}\left(\vecx\mid\vecbeta_g,\vecD_g,\vecA_g,\vecmu_g\right)$ is the density of the MSSAL distribution given in~\eqref{eq:MSSAL} with $\vecbeta_g\in\mathbb{R}^p$, component eigenvector matrix $\vecD_g$, component eigenvalue matrix $\vecA_g$, and component location parameter $\vecmu_g$. 
%This mixture gives components whose contours have the shape of hypercubes (see Figure~\ref{ContCrabs2}, Section~\ref{sec:Crabs2} for example).

\section{Parameter Estimation}\label{sec:PE}

The EM algorithm was formulated in the seminal paper of \cite{dempster77} and is commonly used to estimate the parameters of mixture models. On each iteration of the EM algorithm two steps are preformed: an expectation (E-) step, and a maximization (M-) step. On each E-step the expected value of the complete-data log-likelihood, denoted $\mathcal{Q}$, is calculated based on the current parameter values and on each M-step $\mathcal{Q}$ is maximized with respect to the model parameters. Note that the complete-data refers to the combination of the unobserved (i.e., the missing or latent data) and the observed data. 

For our mixtures of MSSAL distributions, the complete-data are composed of the observed data $\vecx_1,\dots,\vecx_n$, the latent $\vecw_{ig}$, and the component indicators $\vecz_1,\dots,\vecz_n$. Note that for each $i$ and $g$, $\vecw_{ig} = (w_{i1g},\dots,w_{ipg})$ comprise the diagonal elements of the multidimensional weight variable $\vecDelta_{\vecw ig}$, i.e., $$\vecDelta_{\vecw ig} = \diag(w_{i1g},\dots,w_{ipg})$$ for $i=1,\dots,n$ and $g=1,\dots,G$. Furthermore, for each $i$, $\vecz_i=z_{i1},\dots,z_{iG}$ such that $z_{ig} = 1$ if observation $i$ is in group $g$ and $z_{ig}=0$ otherwise, for $g=1,\ldots,G$. 

Using the conditional distributions given in \cite{franczak14}, it follows that 
%\be\label{eq:cond1}
$$\vecX_i\mid w_{ijg}, z_{ig} = 1\backsim\mathcal{N}(\vecmu_g+\vecOmeg_{ig}\vecbeta_g,\vecOmeg_{ig}),$$
%\ee 
where $\vecOmeg_{ig} = \vecD_g\vecA_g\vecDelta_{\vecw ig}\vecD_g'$ for $i = 1,\dots,n$, 
%\be\label{eq:cond2}
$W_{ijg} \mid z_{ig}=1\backsim\text{Exp}(1), $
%\ee
and 
%\be\label{eq:cond3}
$W_{ijg} \mid \vecx_i, z_{ig} = 1 \backsim \text{GIG}(d_{jg},b_{ijg}),$
with $d_{jg} = 2 + [\vecA\vecD_g'\vecbeta]_j^2/a_j$ and $b_{ijg} = [\vecD_g'(\vecx_i-\vecmu_g)]_j^2/a_j$, where $[\vecA\vecD_g'\vecbeta]_j$ denotes the $j$th element of $\vecA\vecD_g'\vecbeta$ and $[\vecD_g'(\vecx_i-\vecmu_g)]_j$ and $a_j$ are defined for~\eqref{eq:NMVM4}. Note that $\text{Exp}(1)$ represents the exponential distribution with rate $1$ and $\text{GIG}(a,b)$ denotes the generalized inverse Gaussian distribution with parameters $a$ and $b$. 
The statistical properties of the GIG distribution are well established and thoroughly reviewed in \cite{jorgensen82}. For our purposes, the most useful properties of the GIG distribution are the tractability of its expected values
\beq
\mathbb{E}\left[ X \right] =
\sqrt{\frac{b}{a}} R_{\nu}\left( \sqrt{ab}\right)~\text{ and }~
%\frac{\sqrt{b}K_{\nu+1}\left( \sqrt{ab}\right) }{\sqrt{a}K_{\nu}\left( \sqrt{ab}\right)},\\
\mathbb{E}\left[{1}/{X}\right] = 
\sqrt{\frac{a}{b}} R_{\nu}\left( \sqrt{ab}\right) -\frac{2\nu}{b},
\eeq
%\end{split}\ee
where $R_{\nu}(c):=K_{\nu+1}\left( c\right)/ K_{\nu}\left( c\right)$. %and $K_{\nu+1}$ and $K_{\nu}$ are defined for~\eqref{eq:SALDens}.

Using the conditional distributions given above %in~\ref{eq:cond1}-\ref{eq:cond3} 
we can now formulate the complete-data log-likelihood for the MSSAL mixtures. Formally, the complete-data log-likelihood is given by
\be\begin{split}
l_{\text{c}}
=\lsum\gsum z_{ig}\log\pi_g &+ z_{ig}\log\phi_p\left(\vecx_i\mid\vecmu_g+\vecOmeg_{ig}\vecbeta_g,\vecOmeg_{ig} \right)\\
&\qquad~~+ z_{ig}\log h_\vecW\left(w_{i1g},\dots,w_{ipg}\right),
\end{split}\ee
where $\pi_g$ are the mixing proportions, $\phi_p\left(\vecx_i\mid\vecmu_g+\vecOmeg_{ig}\vecbeta_g,\vecOmeg_{ig} \right)$ is the density of the multivariate Gaussian distribution with mean $\vecmu_g+\vecOmeg_{ig}\vecbeta_g$ and covariance $\vecOmeg_{ig}$, $h_\vecW\left(w_{i1g},\dots,w_{ipg}\right)=e^{w_{i1g}}\times\dots\times e^{w_{1pg}}$ and $z_{ig}$ and $\vecOmeg_{ig}$ are previously defined.

\subsection{E-step}\label{sec:exp_vals}

For the MSSAL mixtures the expected value of the complete-data log-likelihood on the $(k+1)$th iteration is
\begin{align}
\mathcal{Q} &= \gsum n_g\log{\pi_g}-\frac{np}{2}\log{2\pi} + \gsum\frac{n_g}{2}\log{\left| \bm{\Omega}_{ig}^{-1(k)}\right|} \nonumber\\  
&- \frac{1}{2}\lsum{\gsum{\zig^{(k)}(\vecx_i-\vecmu_g)'\bm{\Omega}_{ig}^{-1(k)}(\vecx_i-\vecmu_g)}}\nonumber\\
&+ \lsum{\gsum{\zig^{(k)}(\vecx_i-\vecmu_g)'\bm{\Omega}_{ig}^{-1(k)}\left(\bm{\Omega}_{ig}^{(k)}\vecbeta_g\right)}}
 - \lsum\gsum\zig^{(k)}\sum_{j=1}^p{E_{1i1g}^{(k)}} \nonumber\\
&+ \frac{1}{2}\lsum{\gsum{\zig^{(k)}\left(\bm{\Omega}_{ig}^{(k)}\vecbeta_g\right)'\bm{\Omega}_{ig}^{-1(k)}\left(\bm{\Omega}_{ig}^{(k)}\vecbeta_g\right)}},\nonumber
\end{align}
where $\bm{\Omega}_{ig}^{(k)} = \vecD_g^{(k)}\bar{\vecDelta}_{\vecw ig}^{(k)}\vecA_g^{(k)}\vecD_g^{(k)'}$, $\bm{\Omega}_{ig}^{-1(k)} = \vecD_g^{(k)}\bar{\vecDelta}_{\vecw ig}^{-1(k)}\vecA_g^{-1(k)}\vecD_g^{(k)'}$, and $\zig^{(k)}$, $\bar{\vecDelta}_{\vecw ig}^{(k)}=\diag\left(E_{1i1g}^{(k)},\dots,E_{1ipg}^{(k)}\right)$, and $\bar{\vecDelta}_{\vecw ig}^{-1(k)}=\diag\left(E_{2i1g}^{(k)},\dots,E_{2ipg}^{(k)}\right)$ are, respectively, the expected values of the sufficient statistics of the component indicators and latent variables. For ease of notation, let $n_g=\lsum\zig^{(k)}/n$. To compute the value of $\mathcal{Q}$ on iteration $(k+1)$, we calculate: 
%On each iteration we require the following expectations for the $E$-step, which effectively amounts to replacing the sufficient statistics of the unobserved data by their expected values. We replace the $z_{ig}$ with their expected value given by
\be\label{eq:zig}
\mathbb{E}[Z_{ig}\mid\vecx_i]=\frac{\pi_g{h}\left(\vecx_i\mid\vecbeta_g^{(k)},\vecD_g^{(k)},\vecA_g^{(k)},\vecmu_g^{(k)}\right)}{\sum_{h=1}^G\pi_h{h}\left(\vecx_i\mid\vecbeta_h^{(k)},\vecD_h^{(k)},\vecA_h^{(k)},\vecmu_h^{(k)}\right)}=:\zig^{(k)},
\ee
%and $\vecDelta_{\vecw ig}$ and $\vecDelta_{\vecw ig}^{-1}$ with the
and let the off-diagonal elements of $\bar{\vecDelta}_{\vecw ig}^{(k)}$ and $\bar{\vecDelta}_{\vecw ig}^{-1(k)}$ be equal to zero and their diagonal elements be equal to
\begin{equation*}\begin{split}\label{eq:MSexp_vals}
&\mathbb{E}[W_{ijg}\mid\vecx_i,z_{ig}=1] = \sqrt{\frac{b_{ijg}}{d_{jg}}} R_{\nu}\left( \sqrt{d_{jg} b_{ijg}}\right)=: E_{1ijg}^{(k)}~\text{ and}\\
&\mathbb{E}[1/W_{ijg}\mid\vecx_i,z_{ig}=1] = \sqrt{\frac{d_{jg}}{b_{ijg}}} R_{\nu}\left( \sqrt{d_{jg}b_{ijg}}\right) -\frac{2\nu}{b_{ijg}}=: E_{2ijg}^{(k)},
\end{split}\end{equation*}
respectively, where $d_{jg} = 2 + [\vecA_g^{(k)}\vecD_g^{'(k)}\vecbeta_g^{(k)}]_j^2/a_j$, $b_{ijg} = [\vecD_g^{'(k)}(\vecx_i-\vecmu_g^{(k)})]_j^2/a_{jg}^{(k)}$, for $j=1,\dots,p$, and $\vecbeta_g^{(k)}$, $\vecD_g^{(k)}$, $\vecA_g^{(k)}$, and $\vecmu_g^{(k)}$ are the values of the model parameters on iteration $(k)$.

\subsection{M-step}

On the $M$-step of the $(k+1)$th iteration the update for $\pi_g^{(k+1)}$ is, in the usual way, given by $\hat{\pi}_g^{(k+1)} = n_g/n$. The updates for $\vecmu_g^{(k+1)}$ and $\vecbeta_g^{(k+1)}$ are given by
\be\begin{split}
\hat{\vecmu}_g^{(k+1)} &= \left(\lsum{\zig^{(k)}\bm{\Omega}_{ig}^{-1(k)}}\right)^{-1}\left(\lsum{\zig^{(k)}\bm{\Omega}_{ig}^{-1(k)}\vecx_i - n_g\vecbeta_g^{(k)} }\right)~\text{and}\\
\hat{\vecbeta}_g^{(k+1)} &= \left(\lsum{\zig^{(k)}\bm{\Omega}_{ig}}^{(k)}\right)^{-1}\left(\lsum{\zig^{(k)}\vecx_i - n_g\vecmu_g^{(k+1)}}\right),
\end{split}\ee
respectively.
To obtain an estimate of $\vecD_g^{(k+1)}$ we employ an iterative optimization procedure. Specifically, the goal is to minimize the function
\be\begin{split}\label{eq:objfunc}
f(\vecD_g^{(k)})
%&=\lsum\tr\left\{\zig^{(k)}\bm{\Omega}_{ig}^{-1(k)}\left(\vecx_i-\vecmu_g^{(k+1)}-\vecOmeg_{ig}^{(k)}\vecbeta_g^{(k+1)}\right)\right.\nonumber\\
%&\qquad\qquad\qquad\left.\times\left(\vecx_i-\vecmu_g^{(k+1)}-\vecOmeg_{ig}^{(k)}\vecbeta_g^{(k+1)}\right)'\right\}\nonumber\\
&=C+\lsum{\tr{\left\{\vecD_g^{(k)}(\bar{\vecDelta}_{ig}^{(k)}\vecA_g^{(k)})^{-1}\vecD_g^{(k)'}\vecW_i^{(k+1)}\right\}}}\\
&\qquad\qquad\qquad\quad - \lsum{\tr{\left\{\vecD_g^{(k)}\bar{\vecDelta}_{ig}^{(k)}\vecA_g^{(k)}\vecD_g^{(k)}\mathbf{B}_i^{(k+1)}\right\}}}
\end{split}\ee
with respect to $\vecD_g$,  where  
\begin{equation*}\begin{split}
\vecW_i^{(k+1)} &= \zig^{(k)}\left(\vecx_i-\vecmu_g^{(k+1)}\right)\left(\vecx_i-\vecmu_g^{(k+1)}\right)',\\ 
\mathbf{B}_i^{(k+1)} &= \zig^{(k)}\vecbeta_g^{(k+1)}\vecbeta_g^{(k+1)'},
\end{split}\end{equation*}
and all expected values and model parameters are previously defined. Typically, the Fleury-Gautschi (FG) algorithm is employed \citep{celeux95,forbes13} but, as noted in \cite{lefkovitch93,boik02,bouveryon07} and \cite{browne12}, the FG algorithm slows down considerably and becomes computationally expensive as the dimension of the data increases. 
%Furthermore, the FG algorithm would not be appropriate for the mixture of MSSALs because the parameter, $\vecbeta_g$, {\color{red}adds an additional term to the objective function} given in~\eqref{eq:objfunc}. 
To circumvent this issue we exploit the convexity of the objective function and construct computationally simpler majorization-minimization (MM) algorithms \citep[see][for examples]{hunter00,hunter04}. Specifically, we follow \cite{kiers02} and \cite{browne13} and preform the following procedure to minimize~\eqref{eq:objfunc}. Note that our MM algorithms use the surrogate function
%\begin{align}\label{eq:surofunc}
$$f(\vecD_g^{(k)})\leq C+\lsum{\tr{\left\{\mathbf{F}_r^{(t)}\vecD_g^{(k)}\right\}}},$$
%\end{align}
where $C$ is a constant that does not depend on $\vecD_g$ and the matrices $\mathbf{F}_r^{(t)}$, for $r = 1, 2$, are explicitly defined below. %in~\eqref{eq:f1} and~\eqref{eq:f2}. 

On iteration $(t)$ compute
\begin{equation*}\begin{split}\label{eq:f1}
\mathbf{F}_1^{(t+1)}&=\lsum{\left[\left(\bar{\vecDelta}_{ig}^{(k)}\vecA_g^{(k)}\right)^{-1} \vecD_g^{'(t)}\vecW_i^{(k+1)}-\omega_{i1}\vecA_g^{-1(k)}\vecD_g^{'(t)}\right]}\\
&\qquad\quad - \lsum{\left[\left(\bar{\vecDelta}_{ig}^{(k)}\vecA_g^{(k)}\right) \vecD_g^{'(t)}\mathbf{B}_i^{(k+1)}-\omega_{i2}\vecA_g^{(k)}\vecD_g^{'(t)}\right]}
\end{split}\end{equation*}
given the current parameter estimates and expected values, where $\omega_{i1}$ and $\omega_{i2}$ are the largest eigenvalues of the matrices $\vecW_i^{(k+1)}$ and $\mathbf{B}_i^{(k+1)}$, respectively. Then calculate the elements of the singular value decomposition of $\mathbf{F}_1^{(t+1)}$, i.e., set
$\mathbf{F}_1^{(t)} =  \mathbf{P}_1\mathbf{B}_1\mathbf{R}_1'$
and find $\mathbf{P}_1$, $\mathbf{B}_1$, and $\mathbf{R}_1'$ where $\mathbf{P}_1$ and $\mathbf{R}_1'$ are orthonormal and $\mathbf{B}_1$ is a diagonal matrix containing the singular values of $\mathbf{F}_1$. It follows that the initial estimate of $\vecD_g^{(k+1)}$ on iteration $(t+1)$ of this optimization procedure is given by
%\be\label{eq:d1}
$\vecD_g^{(k+1)} = \vecD_g^{(t+1)*} = \mathbf{R}_1\mathbf{P}_1'$.
%\ee
 Given this estimate, denoted $\vecD_g^{(t+1)*}$, compute
\begin{equation*}\begin{split}\label{eq:f2}
\mathbf{F}_2^{(t+1)}&=\lsum{\left[\vecW_i^{(k+1)}\vecD_g^{(t+1)*}\left(\bar{\vecDelta}_{ig}^{(k)}\vecA_g^{(k)}\right)^{-1} - \alpha_{i1}\vecW_i^{(k+1)}\vecD_g^{(t+1)*}\right]}\\
&\qquad\quad~~ - \lsum{\left[\mathbf{B}_i^{(k+1)}\vecD_g^{(t+1)*}\left(\bar{\vecDelta}_{ig}^{(k)}\vecA_g^{(k)}\right) - \alpha_{i2}\mathbf{B}_i^{(k+1)}\vecD_g^{(t+1)*}\right]},
\end{split}\end{equation*}
where $\alpha_{i1}$ and $\alpha_{i2}$ are, respectively, the largest eigenvalues of $\left(\bar{\vecDelta}_{ig}^{(k)}\vecA_g^{(k)}\right)^{-1}$ and $\left(\bar{\vecDelta}_{ig}^{(k)}\vecA_g^{(k)}\right)$. Then set $\mathbf{F}_2^{(t+1)}=\mathbf{P}_2\mathbf{B}_2\mathbf{R}_2'$ to obtain 
%\be\label{eq:d2}
$\vecD_g^{(k+1)} = \vecD_g^{(t+1)} = \mathbf{R}_2\mathbf{P}_2'$,
%\ee 
the final estimate of $\vecD_g^{(k+1)}$ on iteration $(k+1)$, where $\mathbf{R}_2$ and $\mathbf{P}_2'$ are also orthonormal. %for~\eqref{eq:d1}.

We repeat the calculations for $\mathbf{F}_1^{(t+1)}$, $\mathbf{F}_2^{(t+1)}$, and $\vecD_g^{(k+1)}$ until the difference in \eqref{eq:objfunc} over consecutive iterations is small. At convergence, we take the final estimate of $\vecD_g^{(t+1)}$ on iteration $(t+1)$ to be the estimate of $\vecD_g^{(k+1)}$.

We maximize $\mathcal{Q}$ with respect to the diagonal matrix $\vecA_g$ via
\be\begin{split}
\vecA_{g}^{(k+1)}& = \diag\left\{\sqrt{\frac{\lsum E_{2i1g}^{(k)} \zig^{(k)} v_{i1g}^{2(k)}}{{n_g}/{a_{1}^{(k)}}+\lsum E_{1i1g}^{(k)}\zig^{(k)}\lambda_1^{2(k)} } },\dots,\right.\\
&\qquad\qquad\qquad~~~\left.\sqrt{ \frac{ \lsum E_{2ipg}^{(k)}\zig^{(k)}v_{ipg}^{2(k)}}{{n_g}/{a_p^{(k)}}+\lsum E_{1ipg}^{(k)}\zig^{(k)}\lambda_p^{2(k)}}}\right\},
\end{split}\ee
where $v_{ijg}^{(k)} = [\vecD_g^{(k+1)'}(\vecx-\vecmu_g^{(k+1)})]_j$, $\lambda_j^{(k)}$ is the $j$th element of the matrix $\load_g=\vecD_g^{(k+1)'}\vecbeta_g^{(k+1)}$, $a_j^{(k)}$ is the $j$th element of the matrix $\vecA_g^{(k)}$, and all off-diagonal elements of $\vecA_g^{(k+1)}$ are equal to zero.

This EM algorithm is considered to have converged when the difference between an asymptotic estimate of the log-likelihood, $l_{\infty}^{(t+1)}$, and the log-likelihood value on iteration $(t)$, $l^{(t)}$, is less than some small value $\epsilon$ \citep{aitken26,bohning94,lindsay95}. %,mcnicholas10a}. 
At convergence we use the final estimates of the $\zig$ to obtain the maximum \textit{a posteriori} (MAP) classification values. Specifically, $\text{MAP}\left\{\zig\right\}=1$ if $\text{max}_h\left\{\hat{z}_{ih}\right\}$ occurs in component $h=g$, and $\text{MAP}\left\{\zig\right\}=0$ otherwise.

\section{Applications}\label{sec:Apps}

\subsection{Model Selection and Performance Assessment}

The Bayesian information criterion \citep[BIC;][]{schwarz78} is used to select the best fitting MSSAL mixture. The BIC was derived via a Laplace approximation and its precision is influenced by the specific form of the model parameters prior densities and by the correlation structure between observations. The BIC is given by 
%\be
$\text{BIC}=2l(\vecx\mid\hat{\varthet})-\rho\log{n}$,
%\ee 
where $l(\vecx\mid\hat{\varthet})$ is the maximized log-likelihood, $\hat{\varthet}$ is the maximum likelihood estimate of $\varthet$, $\rho$ is the number of free parameters in the model, and $n$ is the number of observations. The BIC is commonly used for Gaussian mixture model selection and has some useful asymptotic properties, for example, as $n\rightarrow\infty$ the BIC is shown to consistently choose the correct model \citep[see][for example]{leroux92,dasgupta98}.

The Rand index \citep{rand71} compares partitions based on pairwise agreements. 
%\beq
%\frac{\text{number of agreements}}{\text{number of agreements + number of disagreements}}.
%\eeq
It takes a value between 0 and 1, where 1 indicates perfect agreement. An unattractive feature of the Rand index is that it has a positive expected value under random classification. 
To correct this, \cite{hubert85} introduced the adjusted Rand index (ARI) to account for chance agreement. The ARI also takes a value of~1 when classification agreement is perfect but has an expected value of 0 under random classification. The ARI can also take negative values and this happens for classifications that are worse than would be expected by chance. \cite{steinley04} gives general properties of the ARI and provides evidence supporting its use for assessing classification performance.

\subsection{Illustrative Example: Leptograpsus Crabs}\label{sec:Crabs2}

\begin{figure}[H]
\centering%
\vspace{-0.175in}
\includegraphics[width=0.95\textwidth,height=3.2in]{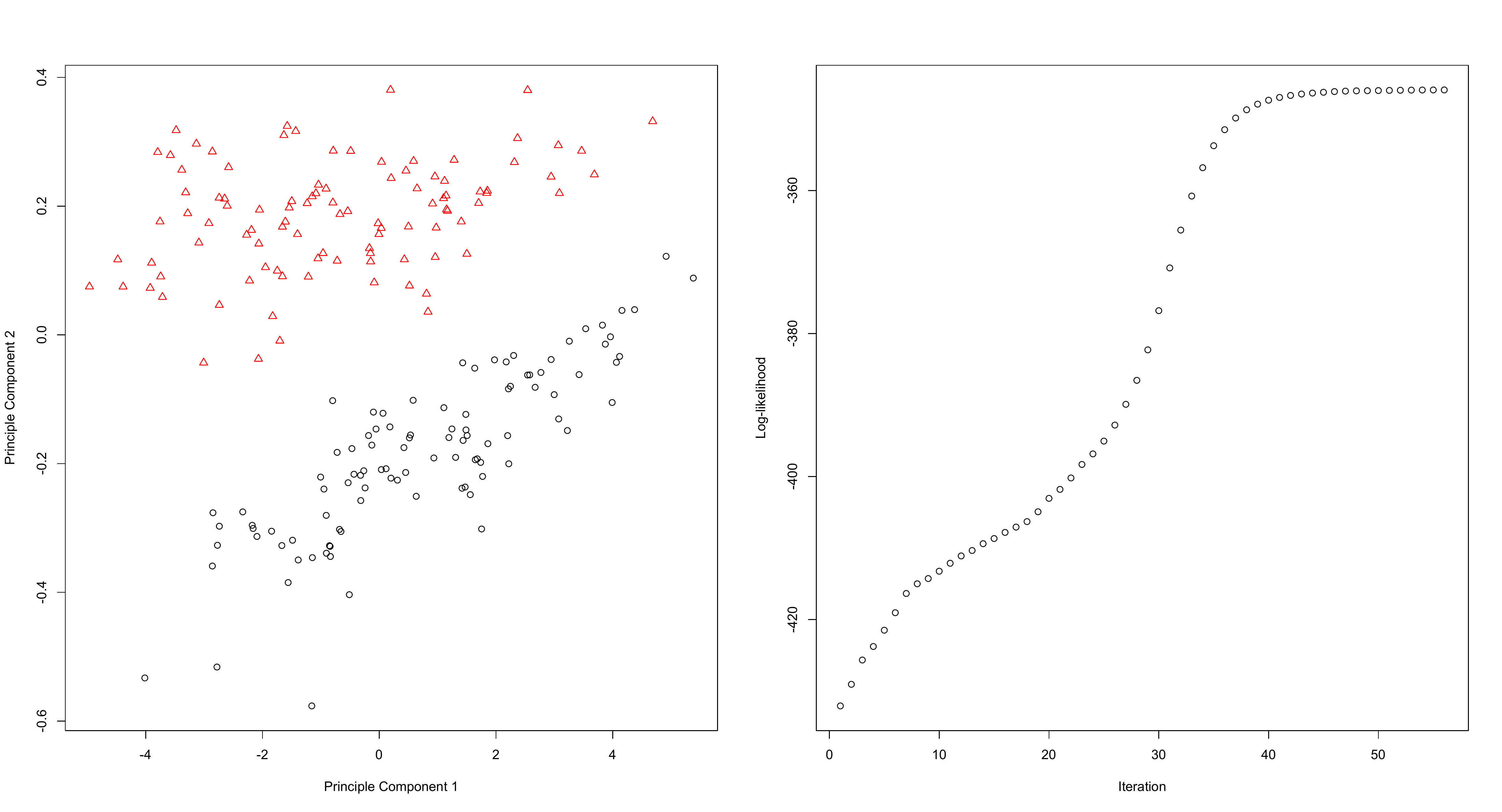}
\vspace{-0.175in}
\caption{The true classifications for the principal components of the Crabs data (Panel 1) and evolution of the associated log-likelihood (Panel 2).\label{CrabsPlot1}}
\end{figure}

\cite{campbell74} give data on 200 crabs of the species \textit{Leptograpsus variegatus} collected at Fremantle, Western Australia. The data are available in the \textsf{R} \citep{R14} package \texttt{MASS} \citep{venables02} and contain five morphological measurements. 
%frontal lobe size (mm), rear width (mm), carapace length (mm), carapace width (mm) and body depth (mm). 
Not surprisingly, the variables in the crabs data are highly correlated with one another and analysis of the covariance matrix using principal components reveals two clusters corresponding to the gender of the crabs (see Panel 1 of Figure~\ref{CrabsPlot1}). 

For strictly illustrative purposes, we remove all group labels and fitted $G=1,\dots,5$ component MSSAL mixtures to the first and third principal components of the \textit{Leptograpsus} crabs data set. Note that for this and each application herein, our MSSAL mixtures are initialized using 50 random starting values. 

\begin{figure}[!ht]
\centering%
\vspace{-0.175in}
\includegraphics[width=0.8\textwidth,height=4in]{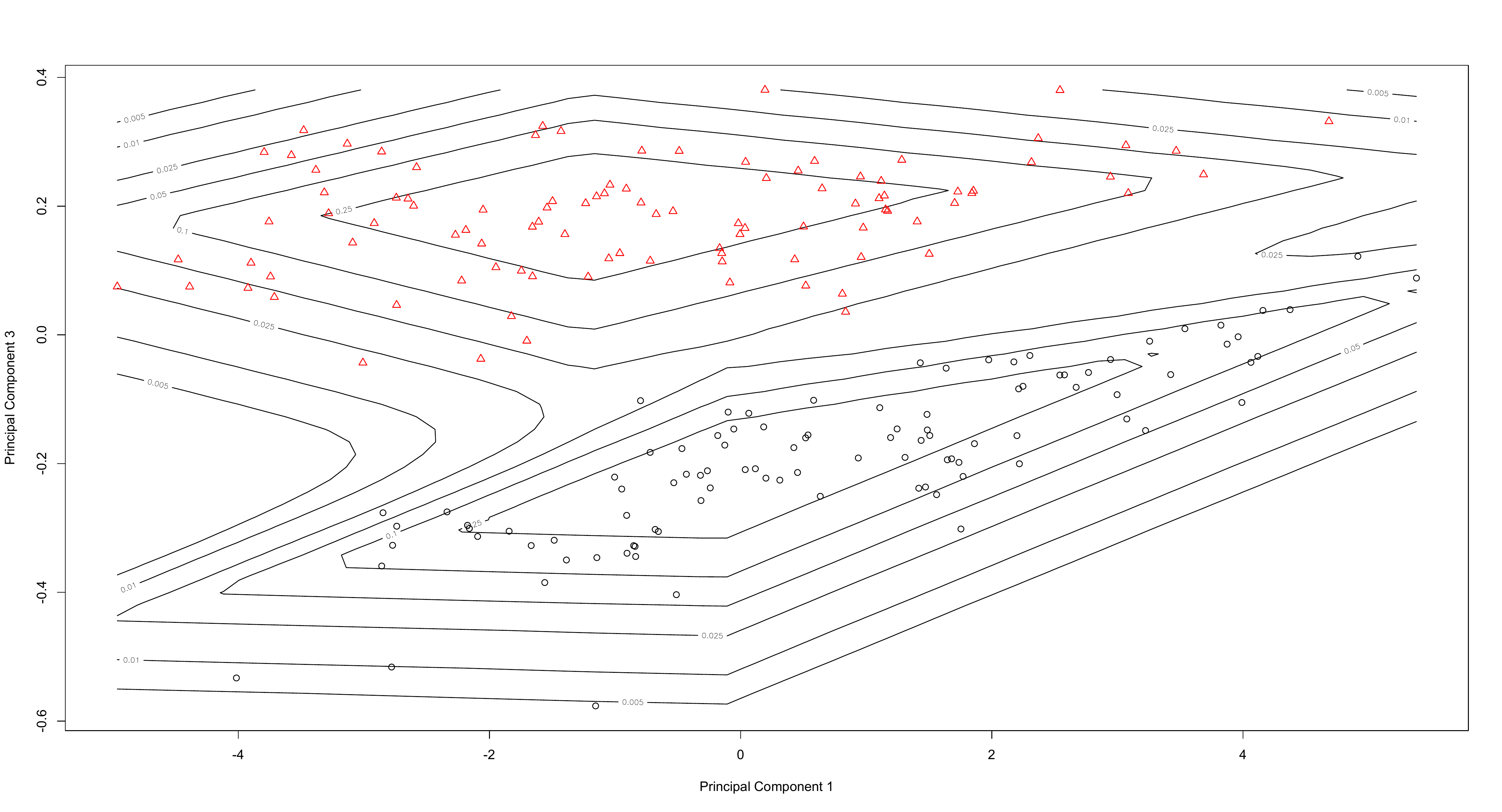}
\vspace{-0.175in}
\caption{The first and third principal components of the crabs data with contours reprinting the fit of the chosen MSSAL model.\label{ContCrabs2}}
\end{figure}

The BIC (-771.3386) selects a $G=2$ component model where the MAP classifications correspond perfectly to gender (i.e., male or female; $\text{ARI}=1.00$). The contour plot for this model (Figure~\ref{ContCrabs2}) shows the unique skewed hypercube shapes of this multivariate generalization. Panel 2 of Figure~\ref{CrabsPlot1} shows the path of the log-likelihood values obtained for the best fitting model on 56 iterations until convergence.

\subsection{Computational Cost}\label{sec:CE}

\begin{figure}
\centering%
\vspace{-0.175in}
\includegraphics[width=\textwidth,height=3.25in]{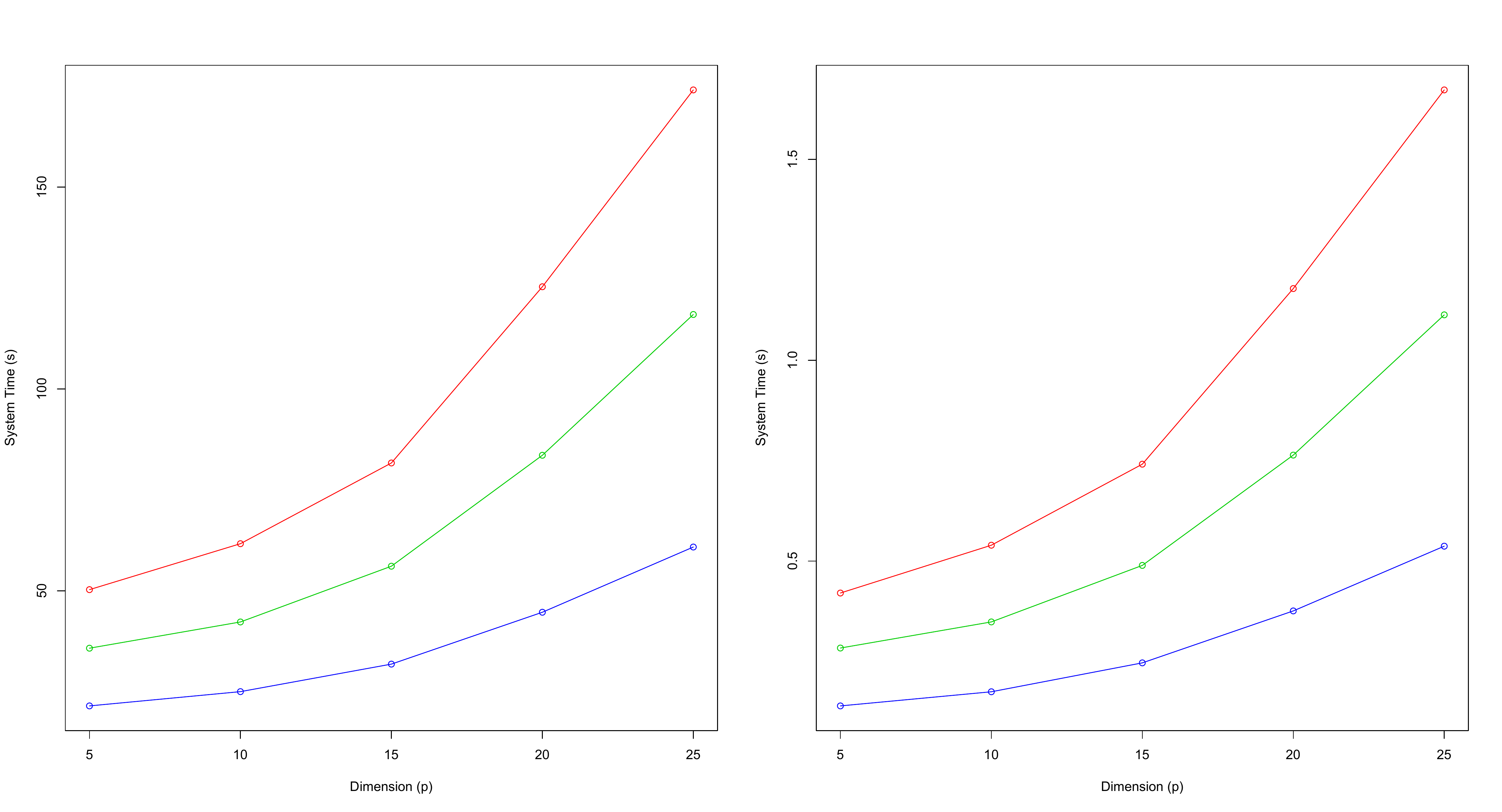}
\vspace{-0.175in}
\caption{Comparison of one, two and three component MSSAL mixtures on the wine dataset. Panel 1 gives the average system time (in seconds) for 100 iterations of our EM algorithm and Panel 2 gives average system time (in seconds) of 1 EM iteration. \label{SysTimes}}
%\caption{Comparison of a one-component and two-component MSSAL mixture on a subset of 50 observations from the golub dataset. \label{SysTimes}}
\end{figure}

In this application we evaluate the speed of our EM algorithm. Specifically, we measure how long it takes to complete one and one hundred iterations of the proposed parameter estimation scheme using one-component, two-component and three-component MSSAL mixtures. We fitted each mixture to subsets of the 27-variable wine data set, available in the \textsf{R} package \texttt{pgmm} \citep{mcnicholas11}. In total there were 5 subsets with $p = 5, 10, 15, 20,~\text{and}~25$ variables, respectively. Panel 1 of Figure~\ref{SysTimes} shows the average elapsed times, in seconds, for the one-component (blue), two-component (green) and three-component (red) MSSAL mixtures to complete one hundred EM iterations. Panel 2 shows the average elapsed time, in seconds, for each MSSAL mixture to complete one EM iteration. 

As expected, the elapsed system time increases with the number of dimensions. Notably, the MSSAL mixtures appear to scale well with dimension as it takes, on average, 61 seconds for our EM algorithm to complete 100 iterations when $G=1$, 118 seconds when $G=2$, and 174 seconds when $G=3$. 

\subsection{Simulation Study: Classification Performance}\label{sec:simu_study}

We use a simulation study to evaluate the classification ability of our MSSAL mixtures. Specifically, we investigate how the mixtures of MSSAL distributions handle symmetric data, skewed Gaussian data, and data generated from a mixture of MSSAL distributions. 

In total, we generate 75 bivariate data sets: 25 from a two-component Gaussian distribution (Scenario I), 25 from a two-component skew-normal distribution (Scenario II) and 25 from a two-component MSSAL distribution (Scenario III). Row 1 of Figure~S.1 (Supplementary Material) %\ref{SimuData1} 
displays the typical shapes from each scenario. We expect good classification performances from each of the models as the data have very little overlap. 

Table~\ref{Simu1} gives the average ARI values, with standard deviations in parenthesis, obtained for the best fitting MSSAL mixtures. We compare these values to the average ARI values obtained from the best fitting mixtures of multivariate SAL, multivariate restricted skew-normal (MSN) and skew-$t$ (MST) and multivariate Gaussian distributions, as chosen by the BIC. Note: the Gaussian mixture models (GMM) were fitted using the \textsf{R} package \texttt{mixture} \citep{browne13a} and the skew-normal and skew-$t$ mixtures were fitted using the \texttt{EMMIXskew} package \citep{mclachlan14}. For this, and the subsequent real data application, all approaches are initialized using 50 random starting values and we remove all group labels. For each scenario the mixtures were fitted for $G=1,\dots,3$ components.
%\begin{figure*}[!ht]
%\centering%
%\vspace{-0.175in}
%\includegraphics[width=\textwidth,height=4.75in]{SimuPlots.pdf}
%\vspace{-0.175in}
%\caption{Row one displays the true groupings and shapes of representative two-component Gaussian, skew-normal, and MSSAL data sets, respectively. Row 2 displays the typical Gaussian (Panel 1 and 2) and skew-normal (Panel 3) solutions when applied to skew-normal  and MSSAL data. \label{SimuData1}}
%\end{figure*}
\begin{table*}[!t]
\centering
\caption{Average ARI values, with standard deviation in parenthesis, for each mixture fitted to data simulated for each scenario. \label{Simu1}}
\vspace{.1in}
\begin{tabular*}{\textwidth}{@{\extracolsep{\fill}}l|lll}
\hline
 & Scenario I & Scenario II & Scenario III \\
\hline
MSSAL & $0.960$  ($0.029$) & $0.932$ ($0.113$) & $\textbf{0.995}$ ($0.006$) \\
SAL & $\mathbf{0.973}$ ($0.016$) & $0.994$ ($0.008$) & $0.413$ ($0.237$) \\
MST & $0.958$ ($0.044$) & $0.994$ ($0.007$) & $0.007$ ($0.013$) \\
MSN & $0.923$ ($0.080$) & $\textbf{0.995}$ ($0.008$) & $0.010$ ($0.020$) \\
GMM & $0.950$ ($0.089$) & $0.976$ ($0.067$) & $0.004$  ($0.006$) \\
\hline
\end{tabular*} 
\end{table*}

\begin{figure}[!t]
\centering%
\vspace{-0.175in}
\includegraphics[width=\textwidth,height=5in]{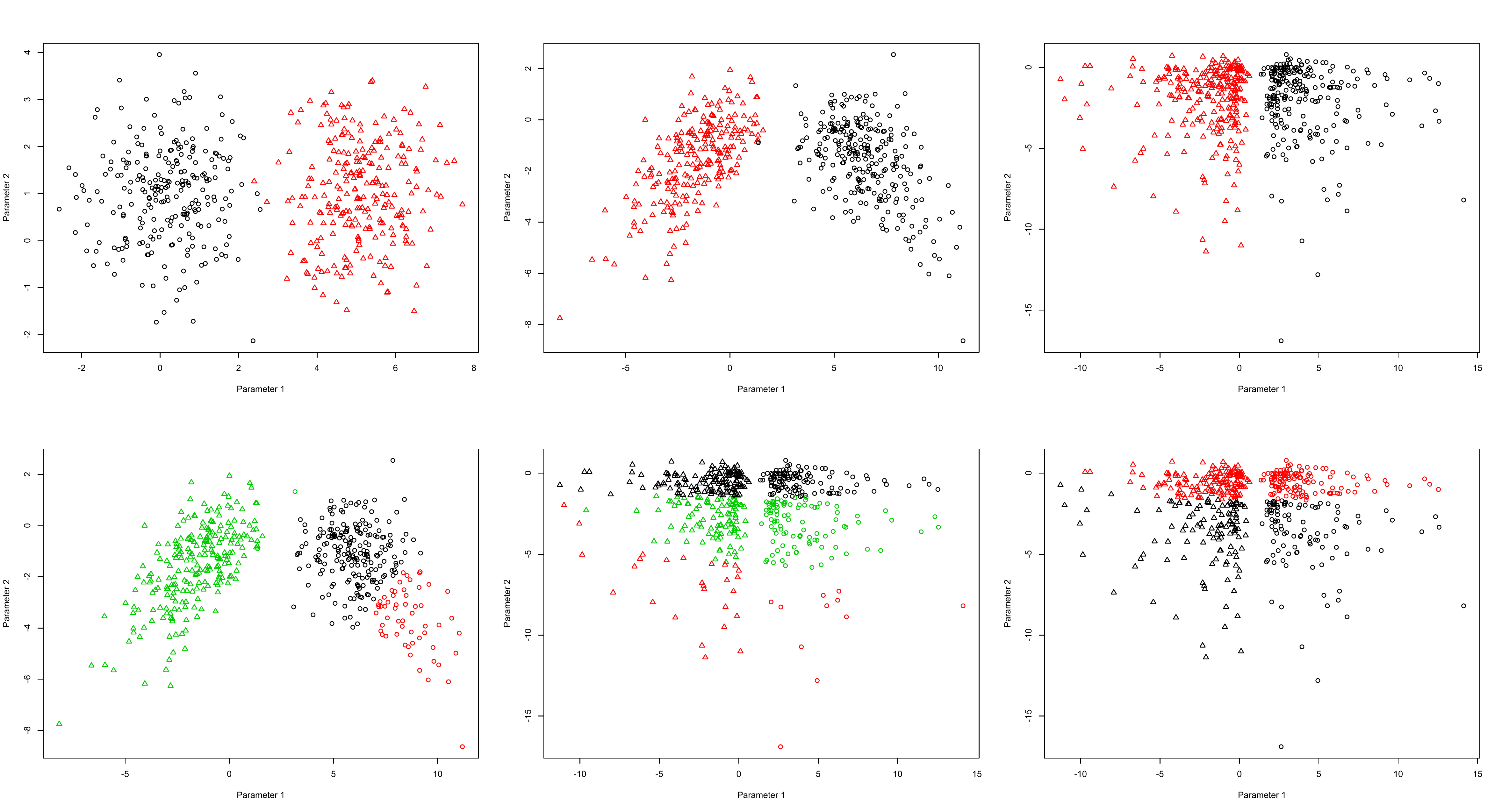}
\vspace{-0.175in}
\caption{Row one displays the true groupings and shapes of representative two-component Gaussian, skew-normal, and MSSAL data sets, respectively. Row 2 displays the typical Gaussian (Panel 1 and 2) and skew-normal (Panel 3) solutions when applied to skew-normal  and MSSAL data. \label{SimuData1}}
\end{figure}

The chosen mixtures of MSSAL distributions give excellent classification results in all three scenarios. For the MSSAL mixtures the BIC chooses the correct number of components $100\%$ of the time in Scenarios I and III and $76\%$ of the time in Scenario II. In Scenario III, the chosen MSSAL mixtures outperform the other mixtures by a substantial margin. With the exception of the mixtures of SAL distributions (where the BIC chooses $G=2$ components for 24/25 data sets) the MST, MSN and GMM mixtures return average ARI values that are essentially no better than random classification. In Row 2 of Figure~S.1, %\ref{SimuData1}, 
the 2nd and 3rd panels show the typical fits of the most popular Gaussian and skew-normal mixtures. It is clear that merging components \citep[see][]{baudry10,hennig10} would not be able to rectify the Gaussian solution. 

Interestingly, the mixtures of multivariate Gaussian distributions gave very good performance on the skew-normal data without the benefit of component merging. The BIC selected $G=2$ component mixtures for $80\%$ of the simulated skew-normal data sets with the other chosen mixtures having three-components. Panel 1 in row 2 of Figure~\ref{SimuData1} displays the typical three-component Gaussian solution for the skew-normal data. It is clear this solution would benefit from merging to give a two component mixture with $\text{ARI} = 1.00$.

Interestingly, the mixtures of multivariate Gaussian distributions fitted the skew-normal data quite well. Specifically, the BIC selects a two-component mixture for $80\%$ of the simulated skew-normal data sets. For the other skew-normal data sets the BIC chooses three-component mixtures. Panel 1 in row 2 of Figure~S.1 %\ref{SimuData1} 
displays the typical three-component Gaussian solution for the skew-normal data. Not surprisingly, this solution could be merged to give a two component mixture with $\text{ARI} = 1.00$.

\subsection{Swiss Banknotes}\label{sec:SwissBank}

The Swiss banknotes data \citep{flury88} are available in the \textsf{R} packages \texttt{alr3} and \texttt{gclus}. In total there are six physical measurements for 100 counterfeit and 100 genuine banknotes. Our goal is to differentiate between each type of banknote. We fitted the mixtures considered in Section~\ref{sec:simu_study} for $G=1,\dots,5$ components using the initialization procedure previously described. Table~\ref{tab:sum_bank} summarizes the performance of the best fitting mixtures.
 \begin{table*}[!t]
\centering%
\caption{Summary of results obtained from the best fitting mixtures fitted to the Swiss banknotes data.}\label{tab:sum_bank}
\begin{tabular*}{\textwidth}{@{\extracolsep{\fill}}l|ccccc}
\hline
Model & MSSAL & SAL & MST & MSN & GMM \\
\hline
\hline
Components & $2$ & $1$ & $2$ & $2$ & $3$  \\
ARI & $\textbf{0.98}$ & $0$ & $0.687$ & $0.687$ & $0.767$ \\
BIC & $-3177.180 $ & $-2900.253$ & $-2845.624$ & $-2734.083$ & $-2740.709$ \\
\hline
\end{tabular*} 
\end{table*} 

\begin{table*}[!t]
\centering%
\caption{A cross-tabulation of true labels and predicted MAP classification results (A, B) for the mixture of MSSAL, MSN and Gaussian distributions, respectively, for the Swiss banknotes data.}\label{ClassBank}
\begin{tabular*}{\textwidth}{@{\extracolsep{\fill}}lccccccccc}
\hline
& \multicolumn{2}{c}{MSSAL} && \multicolumn{2}{c}{MSN} && \multicolumn{3}{c}{GMM} \\
\cline{2-3}\cline{5-6}\cline{8-10}
& A & B && A & B && A & B & C \\
\hline
Genuine & $99$ & $1$ && $83$ & $0$ && $91$ & $9$ & $0$  \\
Counterfeit & $0$ & $100$ && $17$ & $100$ && $0$ & $16$ & $84$ \\
\hline
\end{tabular*} 
\end{table*} 

The results show that the best fitting MSSAL mixture does an excellent job at discriminating between the counterfeit and genuine banknotes, misclassifying only one genuine banknote. On the other hand, both the multivariate skew-normal and skew-$t$ mixtures identified the correct number of groups but misclassified 17 banknotes, the chosen multivariate SAL mixture fail to identify any group structure in this data, and the best fitting GMM uses an extra component. Interestingly, merging Gaussian components would not benefit this solution. Table~\ref{ClassBank} gives the classifications of best fitting MSSAL, MSN (which is identical to the MST) and Gaussian mixtures.

\section{Summary}\label{sec:discussion}

A mixture of MSSAL distributions was introduced and gives mixtures with components whose contours are skewed hypercubes.
Crucially, the level sets of our MSSAL density are guaranteed to be convex, making mixtures thereof ideal for unsupervised learning applications. 
Specifically, the MSSAL distribution is guaranteed to have convex level sets, similar to elliptical distributions like the Gaussian, because it has the same concentration in each direction from the mode. In contrast, the contours for the multiple scaled multivariate $t$-distribution will have levels that are not convex; therefore, situations will arise where one component is used to model two clusters, e.g., X-shaped components.
In addition to the distinct advantage of having convex level sets, our MSSAL distribution has great modelling flexibility. For example, consider the contour plot for the crabs data (Figure~\ref{ContCrabs2}) --- the diamond-like shapes illustrated are a far cry from any of the spherical or tear-drop like densities commonly displayed in the mainstream non-elliptical clustering literature.
In addition to its suitability for unsupervised learning, our MSSAL mixtures should also perform well for supervised and semi-supervised learning, and this will be a focus of future work. Another subject of future work will be exploring evolutionary computation as an alternative to the EM algorithm for parameter estimation, cf.\ \cite{andrews13b} for related work on Gaussian mixtures.

\section*{Acknowledgements}
This work was supported by an Ontario Graduate Scholarship (Franczak), the University Research Chair in Computational Statistics (McNicholas), and an Early Researcher Award from the Ontario Ministry of Research and Innovation (McNicholas).\\

\textbf{This work is currently under consideration at Pattern Recognition Letters}

\bibliographystyle{chicago}
\bibliography{laplace}

\end{document}